\documentclass[sigconf]{acmart}
\usepackage{xcolor}
\usepackage{amsthm}
\usepackage{subcaption}
\usepackage{multirow}
\usepackage{color}
\usepackage{algorithm}
\AtBeginDocument{%
  \providecommand\BibTeX{{%
    \normalfont B\kern-0.5em{\scshape i\kern-0.25em b}\kern-0.8em\TeX}}}

\copyrightyear{2023} 
\acmYear{2023} 
\setcopyright{acmlicensed}\acmConference[WWW '23]{Proceedings of the ACM Web Conference 2023}{May 1--5, 2023}{Austin, TX, USA}
\acmBooktitle{Proceedings of the ACM Web Conference 2023 (WWW '23), May 1--5, 2023, Austin, TX, USA}
\acmPrice{15.00}
\acmDOI{10.1145/3543507.3583505}
\acmISBN{978-1-4503-9416-1/23/04}

\begin{document}

\title{CitationSum: Citation-aware Graph Contrastive Learning for Scientific Paper Summarization}
% Author information can be set in various styles:
% For several authors from the same institution:
% \author{Author 1 \and ... \and Author n \\
%         Address line \\ ... \\ Address line}
% if the names do not fit well on one line use
%         Author 1 \\ {\bf Author 2} \\ ... \\ {\bf Author n} \\
% For authors from different institutions:
% \author{Author 1 \\ Address line \\  ... \\ Address line
%         \And  ... \And
%         Author n \\ Address line \\ ... \\ Address line}
% To start a seperate ``row'' of authors use \AND, as in
% \author{Author 1 \\ Address line \\  ... \\ Address line
%         \AND
%         Author 2 \\ Address line \\ ... \\ Address line \And
%         Author 3 \\ Address line \\ ... \\ Address line}
\author{Zheheng Luo}
\email{zheheng.luo@postgrad.manchester.ac.uk}
\affiliation{%
  \institution{University of Manchester}
  \streetaddress{Oxford Road}
  \city{Manchester}
  \country{UK}
  \postcode{M13 9PJ}
}

\author{Qianqian Xie}
\authornote{Corresponding Author}
\email{qianqian.xie@manchester.ac.uk}
\affiliation{%
  \institution{University of Manchester}
  \streetaddress{Oxford Road}
  \city{Manchester}
  \country{UK}
  \postcode{M13 9PL}
}
\author{Sophia Ananiadou}
\email{Sophia.Ananiadou@manchester.ac.uk}
\affiliation{%
  \institution{University of Manchester}
  \streetaddress{Oxford Road}
  \city{Manchester}
  \country{UK}
  \postcode{M13 9PL}
 }
 
\renewcommand{\shortauthors}{Trovato and Tobin, et al.}

\begin{abstract}
Citation graphs can be helpful in generating high-quality summaries of scientific papers, where references of a scientific paper and their correlations can provide additional knowledge for contextualising its background and main contributions.
Despite the promising contributions of citation graphs, it is still challenging to incorporate them into summarization tasks. This is due to the difficulty of accurately identifying and leveraging relevant content in references for a source paper, as well as capturing their correlations of different intensities.
Existing methods either ignore references or utilize only abstracts indiscriminately from them, failing to tackle the challenge mentioned above.
To fill that gap, we propose a novel citation-aware scientific paper summarization framework based on the citation graph, able to accurately locate and incorporate the salient contents from references, as well as capture varying relevance between source papers and their references. 
Specifically, we first build a domain-specific dataset PubMedCite with about 192K biomedical scientific papers and a large citation graph preserving 917K citation relationships between them.
It is characterized by preserving the salient contents extracted from full texts of references, and the weighted correlation between the salient contents of references and the source paper.
Based on it, we design a self-supervised citation-aware summarization framework (CitationSum) with graph contrastive learning, which boosts the summarization generation by efficiently fusing the salient information in references with source paper contents under the guidance of their correlations.
Experimental results show that our model outperforms the state-of-the-art methods, due to efficiently leveraging the information of references and citation correlations.
\end{abstract}

\begin{CCSXML}
<ccs2012>
<concept>
<concept_id>10002951.10003317.10003347.10003357</concept_id>
<concept_desc>Information systems~Summarization</concept_desc>
<concept_significance>500</concept_significance>
</concept>
</ccs2012>
<ccs2012>
<concept>
<concept_id>10010147.10010178.10010179.10010182</concept_id>
<concept_desc>Computing methodologies~Natural language generation</concept_desc>
<concept_significance>500</concept_significance>
</concept>
<concept>
<concept_id>10010147.10010178.10010179.10010186</concept_id>
<concept_desc>Computing methodologies~Language resources</concept_desc>
<concept_significance>500</concept_significance>
</concept>
<concept>
<concept_id>10010405.10010444.10010450</concept_id>
<concept_desc>Applied computing~Bioinformatics</concept_desc>
<concept_significance>500</concept_significance>
</concept>
</ccs2012>
\end{CCSXML}

\ccsdesc[500]{Information systems~Summarization}
\ccsdesc[500]{Computing methodologies~Natural language generation}
\ccsdesc[500]{Computing methodologies~Language resources}
\ccsdesc[500]{Applied computing~Bioinformatics}

\keywords{Text summarization, citation graph, graph contrastive learning, scientific paper}

\maketitle

\section{Introduction}
\label{sec:intro}
Digital scientific documents such as scientific papers are vital linked web resource.
The rapid growth of scientific research~\cite{10.1371/journal.pone.0093949} requires the development of methods to automatically summarise scientific papers.
Unlike general language texts, scientific papers are characterized by domain specific structures, and domain-specific terms, which must be taken into account when generating succinct and conclusive summaries~\cite{qazvinian2008scientific,abu2011coherent,yasunaga2019scisummnet}.
Furthermore, scientific documents are connected, and relevant to the papers they cite.
The source scientific papers, their references, and their citation correlations, form the enormous citation graph.
In a citation graph, references of scientific papers and their correlations, provide extra knowledge such as the context of its research background, methods, and findings~\cite{cohan2015scientific,qazvinian2008scientific,an2021enhancing,xie2021graph}. 
Therefore, to understand and summarize the gist of a scientific paper, it is essential to incorporate information of its references and relation structure from the citation graph besides the contents of itself.

Although the citation graph plays a promising role in improving the automatic summarization of scientific papers, little attention has been focused on incorporating them in existing pre-trained language models (PLMs) based summarization methods~\cite{liu2019text,lewis2020bart}.
The only exception is CGSum~\cite{an2021enhancing} which leverages references of source papers by constructing a citation graph to improve its summary generation.%for citation graph based summarization method. %with the aim of using references of the source paper to improve its summary generation.
% To leverage the information within references, a citation graph based summarization method CGSum~\cite{an2021enhancing} was firstly proposed to incorporate abstract of papers cited by the source documents.
%It utilized the graph neural networks (GNNs)~\cite{KipfW17} to captures citation relationships, and incorporate abstracts of references with document representations of source papers.
Yet it only incorporates abstracts of references, which can be uninformative or even meaningless to the source document.
As shown in Table \ref{tab:example}, the two abstracts of references have low semantic similarity with the gold summary of the source document (abstract), which shows that abstracts from references can be uninformative for improving the summary generation of the source document.

A better alternative is to incorporate the full contents of references instead of their abstracts. The challenge lies in bringing the full text of the references can introduce redundant information, the content of the references may be irrelevant to the source paper except for some key sentences as shown in Table \ref{tab:related}.
Moreover, different parts of references, such as introduction, related work, methods and experiments, have varying levels of semantic similarities with the source paper. For example, Table \ref{tab:rouge} shows the mean semantic similarity between source documents and different parts of their references in the dataset SSN~\cite{an2021enhancing}.
We observe that sentences selected from the full contents of references have the highest semantic similarity with source documents, while abstracts of references have the lowest semantic similarity.
Therefore, to efficiently leverage references and their correlation structure of the citation graph to improve the summarization of the source document, a remaining big challenge is to identify and locate key information from the full contents of references for the source paper, as well as capture their correlations of different intensities.
This can be laborious even for experienced researchers.
\begin{table}
    \centering
    \footnotesize
    \Description{An example of source document and abstracts of its references in the SSN dataset. We calculate the ROUGE-1 score as the semantic similarity between the gold summary of the source paper (its abstract) and abstracts of its references.}
    \caption{An example of source document and abstracts of its references in the SSN dataset~\cite{an2021enhancing}. We calculate the ROUGE-1 score~\cite{lin2003automatic} as the semantic similarity between the gold summary of the source paper (its abstract) and abstracts of its references.}
    \resizebox{0.48\textwidth}{24mm}{
    \begin{tabular}{|p{1.0\linewidth}|}
    \hline
     \textbf{Abstract of source paper:} this paper summarizes the contents of a plenary talk at the pan african congress of mathematics held in rabat in july 2017. we provide a survey of recent results on spectral properties of schr\"odinger operators with singular interactions supported by manifolds of codimension one and of robin billiards with the focus on the geometrically induced discrete spectrum and its asymptotic expansions in term of the model parameters.\\
     \hline
     \textbf{References 1:} we determine accurate asymptotics for the low-lying eigenvalues of the robin laplacian when the robin parameter goes to \$-infty\$. the two first terms in the expansion have been obtained by k. pankrashkin in the \$ 2d\$-case and by k. pankrashkin and n. popoff in higher dimensions. the asymptotics display the influence of the scalar curvature and the splitting between every two consecutive eigenvalues. (\textcolor{blue}{ROUGE-1: 0.1579})\\
    \hline
    \textbf{References 2:} we give a counterexample to the long standing conjecture that the ball maximises the first eigenvalue of the robin eigenvalue problem with negative parameter among domains of the same volume. furthermore , we show that the conjecture holds in two dimensions provided that the boundary parameter is small. this is the first known example within the class of isoperimetric spectral problems for the first eigenvalue of the laplacian where the ball is not an optimiser. (\textcolor{blue}{ROUGE-1: 0.1818})\\
    \hline
    \end{tabular}}   
    \label{tab:example}
\end{table}
\begin{table}
    \centering
    \footnotesize
    \Description{An example of source document and contents of its reference in the SSN dataset.}
    \caption{An example of source document and contents of its reference in the SSN dataset~\cite{an2021enhancing}. The most related contents in the reference is marked with the blue color.}
    \resizebox{0.48\textwidth}{24mm}{
    \begin{tabular}{|p{1.0\linewidth}|}
    \hline
     \textbf{Source paper:} in this paper, the weak galerkin finite element method for second order elliptic problems employing polygonal or polyhedral meshes with arbitrary small edges or faces was analyzed. with the shape regular assumptions, optimal convergence order for \$ h\^1 \$ and \$ l\_2 \$ error estimates were obtained. also element based and edge based error estimates were proved.\\
     \hline
     \textbf{References:} weak galerkin (wg) refers to a finite element technique for partial differential equations in which differential operators are approximated by their weak forms as distributions. \textcolor{blue}{a weak galerkin method was introduced and analyzed for second order elliptic equations based on weak gradients. in this paper , we shall develop a new weak galerkin method for second order elliptic equations formulated as a system of two first order linear equations}, our model problem seeks a flux function inlineform0 and a scalar function inlineform1 defined in an open bounded polygonal or polyhedral domain inlineform2 satisfying displayform0, and the following dirichlet boundary condition displayform0, where inlineform3 is a symmetric, uniformly positive definite matrix on the domain inlineform4. a weak formulation for ( eqref3 ) - ( eqref4 ) seeks inlineform5 and inlineform6, such that displayform0, here inlineform7 is the standard space of square integrable functions on inlineform8 , inlineform9 is the divergence of vector - valued......... \\
    \hline
    \end{tabular}}
    \label{tab:related}
\end{table}
\begin{table}
\scriptsize
    \centering
    \Description{The mean ROUGE-1 F1 and ROUGE-2 F1 score between different contents of source papers and their references. }
    \caption{The mean ROUGE-1 F1 and ROUGE-2 F1 score~\cite{lin2003automatic} between different contents of source papers and their references. Gold and introduction mean the gold summary (abstract) and introduction of the source paper. For each part of references, we select the top-7 sentences with the greedy search algorithm to calculate the ROUGE score with the source paper.}
    \resizebox{0.48\textwidth}{13mm}{
    \begin{tabular}{c|c|c|c|c}
    \hline
    \textbf{Source}& \multicolumn{2}{c|}{\textbf{Gold}}&\multicolumn{2}{c}{\textbf{Introduction}}\\
    \hline
   References& ROUGE-1& ROUGE-2&ROUGE-1& ROUGE-2\\
    \hline
    Abstract&0.2391&0.0443&0.0267&0.0001\\
    \hline
    Introduction&0.3353&0.0689&0.2042&0.03724\\
    \hline
    Methodology&0.4245&0.0801&0.3718&0.0712\\
    \hline
    Experiments&0.4170&0.0756&0.3562&0.0702\\
    \hline
    Full contents& 0.6737&0.1653&0.6902&0.1808\\
    \hline
    \end{tabular}}
    \label{tab:rouge}
\end{table}
%As shown in Figure \ref{fig:relation}, this results in partially modeling of the hierarchical semantic correlations, that token representations of source documents can not be informed by citing papers and their tokens, or token representations of references can not be informed by source documents and their tokens in otherwise.
%ignoring references, their document contents, and their connections to the source document.
%This results in no information propagation between documents and tokens from different documents, and among tokens of all documents, as shown in Figure \ref{fig:relation}.
%This limits their performance on summary generation since it only allows document-level information interaction within the citation graph. Nevertheless, the summary generation process is conducted on token-level representations token by token.
%It is necessary to capture the hierarchical semantic correlations as mentioned before, to better understand the semantics of the source paper and its tokens.
%Therefore, it requires the method to better understand the meaning of tokens, their contexts, and their connection with other tokens and documents.
%Moreover, CGSum only considers abstracts of references, which  %in which we show more detailed analysis.
%The useful content of the source document can reside on various parts of references
%such as introduction, rather than abstract as shown in Table \ref{tab:rouge}.
%See appendix \ref{sec:cs} for further illustrations and Table \ref{tab:example}, \ref{tab:rouge}.

To fill this gap, we propose a novel citation-aware scientific paper summarization framework based on citation graphs, able to accurately locate and incorporate the salient contents from references, as well as capture the varying relevance between source papers and their references. 
we also developed a new dataset for the biomedical domain: PubMedCite with about 192K biomedical scientific papers and a large citation graph preserving 917K citation relationships among them.
Compared with SSN~\cite{an2021enhancing}, the only other existing scientific document summarization dataset with citations, PubMedCite preserves salient contents extracted from the full textual information of references and the weighted correlation between the salient contents of references and the source paper.
Moreover, it has high domain-specific technical nature and long document contents in references making it more challenging for automatic summarization methods.

We then propose a self-supervised citation-aware summarization framework (CitationSum) with graph contrastive learning, which integrates the salient information from references with the source paper contents under the guidance of their correlations to improve summary generation.
We build a hierarchical heterogeneous graph containing nodes of the source paper, references, and their tokens, based on a weighted citation graph.
We design the contrastive learning guided by the hierarchical graph, to align representations of source documents with key contents of their references from PLMs, at both document and token-level.
This allows our method to incorporate the useful information from references with source papers according to their semantic correlation, and fuse information between source documents, references, and their tokens.
We show that our graph contrastive learning can be deemed as an implicit reconstruction of the weighted citation graph and document contents, which is consistent with the phase of the human writing process for scientific papers.
%Therefore, it can improve scientific summary generation %via better leveraging the rich context around the source documents.
%with a better understanding of source documents.
% We propose to first distill high quality contents of references, to make better use of them.
% For references in the citation graph of the source document, we extract informative contents that have high similarity with the source document, from full contexts of references.
% Next, we build the hierarchical heterogeneous graph with nodes of source paper, references and their tokens, and propose the hierarchical graph guided contrastive learning based on it.
% It models inner-document connections of documents and their tokens, and inter-document connections of source documents and their references in the unified graph structure.
% This enables the bi-directional information propagation among tokens of references or the source paper, references or the source paper, and tokens of the source paper or references.
% %We then propose the hierarchical graph guided contrastive learning to align representations of source documents and their references from PLMs, in both document and token-level.
% We show that the hierarchical graph contrastive learning is implicitly reconstruction of citation graphs, and document contents, which is consistent with the human writing process of scientific papers.
% Thus it can help the model better understand source documents to improve summary generation.
Our main contributions are as follows: 
\begin{enumerate}
    \item We propose a novel self-supervised summarization method CitationSum, that incorporates the graph contrastive learning to incorporate key contents of references and inherited semantic correlations of source documents and references, for scientific paper summarization. 
\item We developed a domain-specific dataset PubMedCite containing 192K biomedical scientific papers and a large citation graph preserving 917K citation relationships among them. To the best of our knowledge, this is the first dataset in the biomedical domain for the task.
\item Experimental results empirically demonstrate that our method can efficiently leverage references and capture semantic correlations between source papers and their references, leading to superior performance compared to previous advanced methods.
\end{enumerate}

\section{Related Work}
\subsection{Scientific Paper Summarization}
Online scientific document as a web-based text resource, its automatic summarization has attracted much attention\cite{10.1145/3184558.3186951, 10.1145/3487553.3524652}.
One direction to tackle this problem is citation-assisted summarization, which aims to highlight the main contributions of papers, based on citation sentences from papers that cite the source document.
The earliest attempts at citation-based summarisation~\cite{qazvinian2008scientific,abu2011coherent,cohan2015scientific}, used the sentence clustering and ranking methods such as Lexrank~\cite{erkan2004lexrank}, to select citation sentences of papers, as the summary for the paper referenced by them.
In addition to the cited text span from papers, \citet{yasunaga2019scisummnet} further utilized the abstract of the target paper, and the graph neural networks (GNNs)~\cite{welling2016semi} to encode all input texts.
\citet{zerva2020cited} investigated the advanced pre-trained encoders BERT~\cite{kenton2019bert} to identify and select citation text spans.
However, these methods cannot address if a newly published paper has not been cited by any paper.
To address this gap, \citet{an2021enhancing} recently proposed the citation graph-based summarization task which considers both the contents of target papers and their corresponding references in the citation graph.
However, they only utilized abstracts of references in a shallow manner, which inspires us to fully leverage references and capture the correlations between source papers and their references, via graph contrastive learning.
%Moreover, there were also attempts investigated the advanced pre-trained language models for scientific paper summarization with only contents of target papers.
%\citet{cachola2020tldr} proposed the extreme summarization task, namely TLDR generation for scientific papers, and investigated the BART~\cite{lewis2020bart} based methods.
%\citet{gupta2021effect} analyzed factors that influence the performance of fine-tuning pre-trained language models on scientific document summarization. 
\subsection{Text Summarization with Contrastive Learning}
Recently, contrastive learning has been introduced to improve text summarization.
\citet{liu2021simcls} proposed to use the contrastive learning to optimize the quality of generated summaries according to the evaluation metric ROUGE.
\citet{cao2021cliff} and \citet{nan2021improving} utilized the contrastive learning to improve the factuality of abstractive summarization.
\citet{liu2021topic} proposed the topic-aware contrastive loss to capture dialogue topic information for abstractive dialogue summarization.
\citet{wang2021contrastive} designed the contrastive loss to align sentence representation across different languages, for multilingual summarization.
\citet{hu2022graph} used the contrastive learning to improve the graph encoder, for the radiology findings summarization.
Different from them, we focus on the citation graph-based summarization task, which has been rarely studied.
\citet{xie2022gretel} designed the graph contrastive learning to capture the better topic information for long document summarization. 

\section{Citation-aware PubMed Dataset}
To support the evaluation and development of our method, we first introduce a new large-scale citation-aware PubMed dataset (PubMedCite) with 192,744 scientific paper nodes and 917,838 citation relationships in the biomedical domain, that are extracted from the PubMed Central Open Access Subset \footnote{\url{https://www.ncbi.nlm.nih.gov/pmc/tools/openftlist/}}. %It is noticed that we can't release the dataset directly due to license limitation (such as CC BY-SA, CC BY-ND licenses) of some papers.
The PubMedCite corpus is built on the PubMed Central Open Access Subset.
To construct PubMedCite, we first downloaded the whole PubMed Open Access Subset (up to 17 Nov. 2021), then build the graph by adding papers into it through the breadth first traversal starting from a random document until the number of nodes reaches a limit.
During the construction, we utilize \textit{pubmed parser} by~\citet{Achakulvisut2020} to extract the \textit{PMC id}, \textit{pubmed id}, title, abstract, and full article of each document.
For the inductive setting, we used the same graph building method to sample two different sub-graphs from the whole PubMedCite citation graph as the validation and test set. Then, we removed the inter-graph edges among the three sets to ensure their independence.
Although it is not for commercial use, we are unable to release document contents of the dataset directly due to license limitations (such as CC BY-SA, and CC BY-ND licenses) of some papers.
We will release the build citation graph among all documents, and provide the code script for users to access and process the document contents themselves, according to the paper id saved in the citation graph.

The statistics of the PubMedCite dataset and comparison with the only existing dataset SSN~\cite{an2021enhancing} are shown in Table~\ref{tab:stats}.
\begin{table*}[!hbt]
    \small
    \centering
    \Description{The statistics of two tested datasets}
    \caption{The statistics. "Sum words" and "Sum sent" denote the average word and sent number of summarizations. }
    \begin{tabular}{c|cccccccccccc}
    \hline
    Dataset& Train&Dev&Test&Word&Sent&Sum words&Sum sent&Edges&Domain\\
    \hline
    SSN&128,400&6,123&6,276&5072.3&290.6&165.1&6.4&660,908&General\\
    \hline
    Our&178,100&7,036&7,608&4409.7&159.8&265.1&10.1&917,838&Biomedical\\
    \hline
    \end{tabular}
    \label{tab:stats}
\end{table*}
When compared with SSN: 1) The average length of gold summary in our dataset is longer than that of SSN, while the average length of full articles is relatively shorter than that of SSN.
2) Our dataset keeps full contents of references to make better use of their information, while SSN only keeps abstracts of them.
3) Our dataset only includes papers in the biomedical domain, while SSN consists of papers from several different fields including mathematical, physics, and computer science, our dataset can help the evaluation of domain-specific tasks in the research community.
Moreover, biomedical scientific papers are laden with terminology and have complex syntactic structures~\cite{wang2021pre,luo2022readability}.
This makes our dataset a challenging benchmark for automatic summarization methods.

\section{Methods}
We first define the task of scientific paper summarization with the citation graph.
Given a corpus $D$, each document $d$ in the corpus is represented by the sequence of $n$ tokens: $x = \{x_1,x_2,\cdots,x_n\}$.
Its target summary is represented by a sequence of $m$ tokens: $y = \{y_1,x_2,\cdots,y_m\}$, where $m \ll n$.
The citation graph $G = \{V, E\}$ of the corpus preserves citation relationships among all documents, where $V$ is the set of document nodes and $E$ is the set of edges.
$E$ can be represented by the adjacency matrix $A$, where $A_{d,d'} = 1$ means there exists a citation link between document $d$ and $d'$, $A_{d,d'} = 0$ otherwise.
For each source document $d$, it aims to generate its target summary $y$ based on the source paper and the sub-citation graph $A^d$ with only the source document and all its neighbours. 
It is generally considered as a conditioned sequence-to-sequence~\cite{sutskever2014sequence} learning problem to model the generation process $p(y|x,A^d)$.
\begin{figure}
  \Description{the architecture of our graph contrastive summarisation model}
  \centering
      \includegraphics[width=1\linewidth]{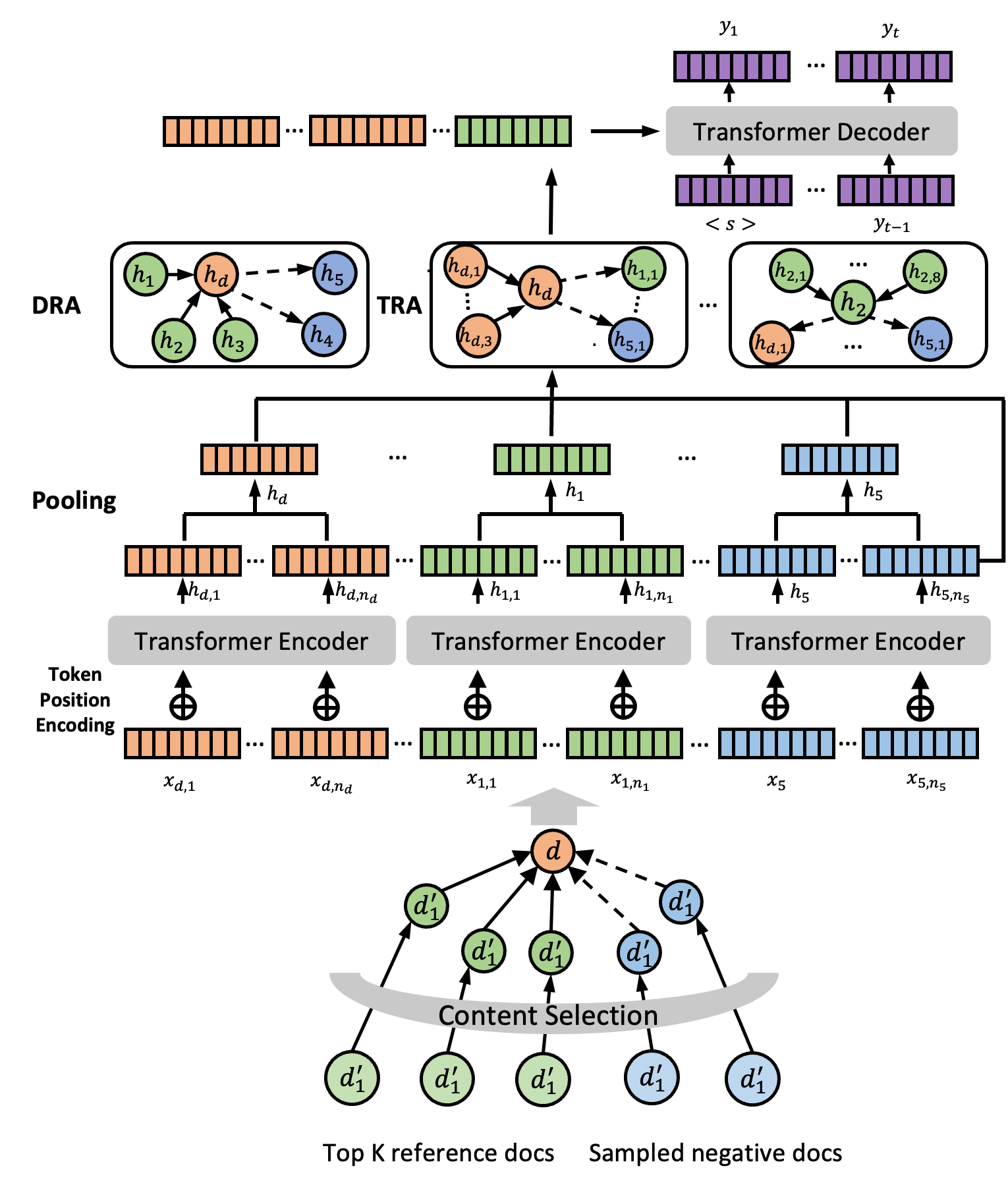}
    %   \captionsetup{font={scriptsize}}
       \caption{The model architecture.}
\label{fig:model}
\end{figure}

In this section, we will introduce our proposed citation-aware scientific paper summarization framework (CitationSum) based on the graph contrastive learning and the citation graph. CitationSum aims to fully leverage the useful information of references and the weighted citation correlation in the citation graph. 
Different from previous methods which used only the abstracts of references ~\cite{an2021enhancing}, we first select the key information from the full contents of references for each source document and build its weighted citation graph to capture the varying semantic correlation between the source paper and its references.
To further make the deep information fusing between the source paper and its references, we build the hierarchical heterogeneous graph based on the weighted citation graph and document contents, to capture the semantic correlations between the source paper, references, and their tokens.
As shown in Figure \ref{fig:model}, we encode the source document and its references with the PLMs-based encoder to yield token-level representations.
We use the pooling strategy to further generate document representations for the source paper and its references, based on the token representations.
The self-supervised graph contrastive learning is designed to align representations of the document and references at both document and token levels, through the hierarchical heterogeneous graph.
This allows the model to integrate the salient information of references into the source paper according to their varying semantic similarity, to improve the summarization generation of the source paper.
Finally, the citation-aware token representations of the source paper are fed into the decoder to conduct the summarization generation.
\subsection{Input Representation}
\textbf{Contents Selection}
As shown in Table \ref{tab:rouge},
sentences that are relevant to the source paper are found in the full contents of references.
Unlike previous methods considering only abstracts~\cite{an2021enhancing}, we first aim to identify the most useful contents of references to make better use of them.
Similar to the oracle summary selection, for each source document, we use a greedy selection algorithm~\cite{nallapati2017summarunner,liu2019text} to extract the top sentences from the full contents of the references. This process maximises the ROUGE score against the target summary of the source document (generally the abstract). 
This heuristic approach iteratively selects one sentence adding to the summary til the ROUGE score cannot be improved by adding any more to the summary.
It is noticed that we use the introduction of the source document instead of the target summary to make the content selection for the test set since the target summary of the source document should be unseen during the test process.
As shown in Table \ref{tab:rouge}, the abstract and introduction of the source paper have comparable semantic similarity with references, demonstrating the necessity of extracting key contents from introductions.

We also narrow down the sub-graph by only considering the most relevant neighbour nodes from the full citation graph $A^d$ for each document $d$, following the previous method~\cite{an2021enhancing}.
To reduce the computational costs of processing all references and encoding the full graph, we sample the sub-graph with the top neighbour references from the full citation graph $A^d$ for each document $d$.
Our method differs from ~\citet{an2021enhancing} that extracted neighbours based on their hidden representations, we select references that have the maximum semantic similarity (measured by the ROUGE score) with the source document from neighbour references.
%sentences selected from full contents of references have the highest semantic similarity with source papers when compared with other contents.
%Therefore, we first aim to select the most useful contents of references to make better use of them, different from the previous method~\cite{an2021enhancing} incorporated their abstracts.
%The unsupervised greedy approach iteratively selects one sentence to the summary, until the ROUGE score has been maximized.
%Although there are previous methods 

\textbf{Input Encoder}. Given the source document $d$ by the sequence of tokens $\{x_1,x_2,\cdots,x_n\}$, 
and its references $d' \in \mathcal{N}_d$ in the sub citation graph $A'_d$, we first convert each token of them into the sum of token embedding, position embedding and segmentation embedding.
We then yield contextual representations of tokens with the pre-trained language model: $h_d\{*\}=\{h_{d,1},h_{d,2},\cdots,h_{d,n}\}=LM(\{x_1,x_2,\cdots,x_n\})$.
Finally, we aggregate token representations with the pooling layer to achieve the contextual representation of source document: $h_d = FFN([max(h_d\{*\}), mean(h_d\{*\})])$, where $FFN$ is the feed-forward network.
Under the same process, we yield the contextual representation of its references $h_{d'} (d' \in \mathcal{N}_d$), where $\mathcal{N}_d$ is the set of neighbours of $d$ in the sub-citation graph.
It is noticed that the source paper and its references are input individually into PLMs.

\subsection{Citation-aware Graph Contrastive Learning}
We leverage information from references and the correlation structure in the citation graph to guide summary generation by proposing the self-supervised citation-aware graph contrastive learning framework. This enables a rich information fusion between source documents and their references. 
%and capture which also allows us to model relationships between them in the citation graph simultaneously.

\subsubsection{Hierarchical Graph Construction}
\label{sub:hgc}
For each source document, we first build a hierarchical heterogeneous graph with multi-granularity nodes.
It consists of a two-level graph organized hierarchically: the weighted citation graph to capture the citation correlations between the source document and its references, and the document graph to model the correlation between documents and tokens.

\textbf{Weighted Citation Graph Construction}. For each source document $d$, and all its selected k-hop neighbour references in its sub-citation graph, we build weighted edges for them according to their semantic similarity.
We also randomly select documents that have no citation relation with the source document, as the negative nodes in the graph.
The edge of two nodes $(i, j)$ is defined as:
\begin{equation}
\small
   \begin{aligned}
{A'}_{i,j}^d = \begin{cases} ROUGE({Abst}_i ,{Cont}_j), & i=d, j \in \mathcal{N}_d\\ 
ROUGE({Cont}_i, {Cont}_j), & j \in \mathcal{N}_d\\
                         1, & i=j\\
                         0, & otherwise
          \end{cases}
\label{eq:a}
\end{aligned}
\end{equation}% \vspace{-0.25cm}
where $ROUGE({Abst}_i, {Cont}_j)$ is the mean score of ROUGE-1 and ROUGE-2 between the abstract of the source document and key contents of its neighbours, $ROUGE({Cont}_i, {Cont}_j)$ is the mean score of ROUGE-1 and ROUGE-2 between highlight contents of neighbour $i$ and $j$. 
Edges with weights ${A'}_{i,j}^d$ that are less than $\rho$, will be deleted to avoid introducing noise.

\textbf{Bipartite Document Graph Construction}. We build the bipartite graph for each document $d$ (including source and references), 
whose nodes are documents and tokens, and edges are occurrences of tokens at the document.
%that captures occurrences of tokens at documents.
We also randomly select tokens from other documents as the negative token nodes in the graph.
The edge of two nodes $(i, j)$ is defined as:
\begin{equation}
\small
   \begin{aligned}
B_{i,j}^d = \begin{cases} 1, & i=d, j \in \{x_1,x_2,\cdots,x_n\}\\ 
                         0, & otherwise
          \end{cases}
\label{eq:bg}
\end{aligned}
\end{equation}
%As shown in Figure \ref{fig:graph}, the hierarchical graph enables the bi-directional message passing among tokens-references-source-tokens, for example $x_{1,1}' \Leftrightarrow d'_1 \Leftrightarrow d \Leftrightarrow x_1$.
%Different from previous methods, %the source document can now be informed by both document and token representations of their neighbours.
%for all document nodes, their representations can be influenced by both neighbour documents and especially tokens of all neighbour documents, apart from their own tokens.
%Similarly, representations of all token nodes can be affected by not only their own documents but also neighbour documents and all tokens of neighbour documents.
% For both the source document and its connected references, we can see that their document representation and token representations are informed by token and document representations of their neighbours.
% Token and document representations of the source document and its references, will be greatly influenced
% %by more semantically similar contents 
% by tokens with higher occurrence frequency 
% in their own document content and neighbours.
% For a reference, it will contribute more to representations of the source document, if it also has correlations to other references of the source document.
%\begin{figure}
%  \centering
%      \includegraphics[width=0.8\linewidth]{graph.pdf}
%    %   \captionsetup{font={scriptsize}}
%       \caption{The hierarchical graph.}
%\label{fig:graph}
%\end{figure}

\subsubsection{Graph Contrastive Learning}
\textbf{Document Representation Alignment}. Based on the weighted citation graph, we design the following document representation alignment (DRA) loss:
\begin{equation}
\small
\mathcal{L}_{DRA} = -\frac{1}{N_d}\sum_{d=1}^{N_d}log(\frac{\sum_{0<A^{\prime d}_{i,j}}-{\hat{A}}^{\prime d}_{i,j} e^{h_i \cdot h_j}}{\sum_{A^{\prime d}_{i,j}=0}e^{h_i \cdot h_j}})
\label{eq:dra}
\end{equation}
where ${\hat{A}}^{\prime d}=I-D^{-\frac{1}{2}}{A'}^d D^{-\frac{1}{2}}$ is the normalized graph Laplacian of $A^{\prime d}$, $D$ is the degree matrix of $A'^d$, $N_d$ is the number of documents, and $h_{i}, h_j$ are document contextual representations from the PLM encoder.
It encourages representations of documents (including the source and its references), and their neighbours to be closer, while pulling away representations of documents that don't have citation correlation.
This makes explicitly information fusion between source documents and key contents of their references according to their semantic correlation in the citation graph, to yield citation-aware document representations for source documents and their neighbours.

\textbf{Token Representation Alignment}. %It is intuitive that correlated documents should be close in the semantic space, or in otherwise be far way.
To further make information propagation between token representations of source documents and their neighbours, we then design the document graph-guided contrastive learning, to align token representations with citation aware-document representations.
We design the following token representation alignment (TRA) loss based on the bipartite graph:
\begin{equation}
\small
\mathcal{L}_{TRA} = -\frac{1}{N_d}\sum_{d=1}^{N_d}log(\frac{\sum_{0<B^d_{d,j}}-{\hat{B}^d}_{d,j} e^{h_d \cdot h_j}}{\sum_{B^d_{d,j}=0} e^{h_d \cdot h_j}})
\label{eq:g}
\end{equation}
where $\hat{B}^d=I-D^{-\frac{1}{2}}{B}^d D^{-\frac{1}{2}}$ is the normalized graph Laplacian of ${B}^d$, and $h_{d}, h_j$ are document and token representations from the PLM encoder.
It pushes the citation-aware representation of the document and representations of tokens closer if these tokens appeared in the document, and pulls away otherwise.

The alignment makes the information of references be propagated into token representations of source documents, and also token representations of references be grounded by representations of source documents.

\subsection{Graph Contrastive as Matrix Factorizing}
In this section, we aim to make a theoretical understanding of the information captured by the hierarchical graph contrastive learning process in the above.
We can find that the hierarchical graph contrastive learning based on the weighted citation graph and document graph, can be reformulated as the implicit matrix factorization, to reconstruct them.

For the equation \ref{eq:g}, we derive its upper bound as:
\begin{equation}
\small
\mathcal{L}_{TRA} \leq -\frac{1}{N_d}\sum_{d=1}^{N_d} log(\frac{\sum_{0< B^d_{d,j}} e^{h_d \cdot h_j}}{\sum_{o=1}^{N_t} e^{h_d \cdot h_o}})
\label{eq:r}
\end{equation}
where $N_t$ is the number of tokens (including positive and negative tokens).
Since $-{\hat{B}^d}_{d,j}=\frac{B^d_{d,j}}{\sqrt{{|N|}_d \cdot {|N|}_j}}=\frac{1}{\sqrt{N_t}}$ is the same value for all positive token $j$, thus can be dependent from the log function and collapsed to a constant, where ${|N|}_d=N_t, {|N|}_j$ are the degree of document node $d$, and token node $j$. 

To minimize the $\mathcal{L}_{TRA}$, we can instead optimize its upper bound.
Following previous methods~\cite{mikolov2013distributed, levy2014neural}, the upper bound can be approximated with the negative sampling:
\begin{equation}
\small
\begin{aligned}
 \max &\frac{1}{N_d}\sum_{i=1}^{N_d}\sum_{j \in \mathcal{N}^+_d}n_{d,j} log(\sigma(h_d \cdot h_j)) +\\
 &k \cdot \mathbf{E}_{o \in \mathcal{N}} log \sigma(-h_d \cdot h_o)
 \end{aligned}
\label{eq:rw}
\end{equation}
where $n_{d,j}$ is the appearance frequency of token $j$ in document $d$, $\mathcal{N}^+_d$ is the set of positive tokens appeared in document $d$, $k$ is the number of sampled negative tokens, $\sigma(x)=\frac{1}{1+e^{-x}}$, and $\mathcal{N}$ is the set of tokens in the corpus.
For simplify the analysis, we consider the negative tokens are sampled from the empirical unigram distribution, thus the expectation term in equation \ref{eq:rw} can be rewritten as:
\begin{equation}
\begin{aligned}
\mathbf{E}_{o \in \mathcal{N}} log \sigma(-h_d \cdot h_o)
&= \sum_{j \in \mathcal{N}^+_d} \frac{n_j}{N_d} log \sigma(h_d \cdot h_j)\\
&+\sum_{o \in \mathcal{N}^-_d} \frac{n_o}{N_d} log \sigma(-h_d \cdot h_o)
 \end{aligned}
\label{eq:ex}
\end{equation}
where $n_j$ is the appearance frequency of the positive token $j$ in the corpus, $n_o$ is the appearance frequency of the negative token $o$ in the corpus.

To optimize the equation \ref{eq:rw}, we consider $x=h_d \cdot h_j$ and yield the its partial derivative with $x$ after explicitly representing the expectation term with equation \ref{eq:ex}:
\begin{equation}
\small
\frac{\partial \ell}{\partial x} = n_{d,j} \sigma(-x) + k \cdot \frac{n_j}{N_d} \sigma(x)
\label{eq:pd}
\end{equation}
where $n_j$ is the appearance frequency of the positive token $j$ in the corpus.
%We set the partial derivative to zero in equation \ref{eq:rw}, and have:
%\begin{equation}
%e^{2x} - (\frac{n_{d,j}}{k \cdot \frac{n_j}{N_d}} - 1) e^x - \frac{n_{d,j}}{k \cdot \frac{n_j}{N_d}} = 0
%\label{eq:de}
%\end{equation}
%To solve the quadratic equation about $e^x$, we can yield:
%\begin{equation}
%\begin{aligned}
%e^x &= \frac{n_{d,j}}{k \cdot \frac{n_j}{N_d}}\\
%x &= log n_{d,j} + log \frac{N_d}{n_j} - log k\\
%h_d \cdot h_j &= log n_{d,j} + log \frac{N_d}{n_j} - log k\\
%\end{aligned}
%\end{equation}
We set the partial derivative to zero, and achieve the following maximum point:
\begin{equation}
\small
h_d \cdot h_j = log n_{d,j} + log \frac{N_d}{n_j} - log k
\label{eq:mp}
\end{equation}
We can see that the optimized $h_d, h_j$ aims to reconstruct the shifted log appearance frequency of token $j$ in document $i$, that is regularized by the inverse document frequency (IDF): $log \frac{N_d}{n_j}$.

For the equation \ref{eq:dra}, we derive its upper bound as:
\begin{equation}
\small
\mathcal{L}_{DRA} \leq -\frac{1}{N_d}\sum_{d=1}^{N_d} log(\frac{\sum_{0< A'^d_{i,j}} -\hat{A}'^d_{i,j} e^{h_i \cdot h_j}}{\sum_{o=1}^{N_g} e^{h_i \cdot h_o}})
\label{eq:ub}
\end{equation}
where $N_g$ is the number of nodes in the citation graph (including positive and negative nodes).
According to the Jensen's inequality~\cite{hansen2003jensen}, we further rewrite the equation \ref{eq:ub} as:
\begin{equation}
\small
\mathcal{L}_{DRA} \leq -\frac{1}{N_d}\sum_{d=1}^{N_d} -\hat{A}'^d_{i,j} log(\frac{\sum_{0< A'^d_{i,j}} e^{h_i \cdot h_j}}{\sum_{o=1}^{N_g} e^{h_i \cdot h_o}})
\label{eq:reee}
\end{equation}
We approximate the upper bound with the negative sampling:
\begin{equation}
\small
\begin{aligned}
 \max &\frac{1}{N_d}\sum_{d=1}^{N_d}\sum_{0< A'^d_{i,j}} -\hat{A}'^d_{i,j} log(\sigma(h_i \cdot h_j)) +\\
 &k \cdot \mathbf{E}_{A'^d_{i,o}=0} log \sigma(-h_i \cdot h_o)
 \end{aligned}
\label{eq:ns}
\end{equation}
It has similar formulation with the equation \ref{eq:rw}, thus we yield its optimized point as following:
\begin{equation}
\small
h_i \cdot h_j = log -{\hat{A}}^{\prime d}_{i,j} + log \frac{N_d}{n_d^+} - log k
\label{eq:ope}
\end{equation}
where $n_d^+$ is the number of neighbours of document $d$.
We can find that optimizing the $\mathcal{L}_{DRA}$ loss implicitly refers to factorizing the shifted log weighted citation matrix $A'^d$.

The implicitly reconstructing of the log citation graph $A'^d$ is similar to the graph auto-encoder~\cite{kipf2016variational}, which can learn representations of source documents efficiently via capturing the topological structure information in the citation graph.
The log appearance frequency of each token is similar to the document topic modeling process~\cite{blei2003latent,xie2021graph,xie2021grtm}, thus capturing the global context semantics of documents.
The analysis helps to explain why and how the designed hierarchical graph contrastive learning on improving representations of source documents from the perspective of reconstructing document contents and their correlation structure.

\subsection{Decoder}
We use the standard transformer~\cite{vaswani2017attention} based decoder similar to previous methods~\cite{liu2019text, see2017get}.
We feed token representations of document $d$ and its references, along with previously generated tokens $y_{t-1}^d$ to get the current output:
\begin{equation}
\small
\begin{aligned}
h_t^d &= \text{LN}(y_{t-1}^d+\text{SELFATTN}(y_{t-1}^d))\\
h_t^d &= \text{LN}(h_t^d+\text{CROSSATTN}(h_t^d, \hat{h}_d))\\
h_t^d &= \text{LN}(h_t^d+\text{FEEDFORWARD}(h_t^d))\\
\end{aligned}
\label{eq:deco}
\end{equation}
where $\hat{h}_d$ is the set of token representations of document $d$ and its references based on the graph contrastive learning, \text{LN} means layer normalization, and $h_t^d$ is the hidden representation of the current token in the decoder.
The final training loss is:
\begin{equation}
\small
\mathcal{L} = -\frac{1}{N_d}\sum_{d=1}^{N_d} \sum_{t=1}^n p(y_t^d|\hat{y}_{<t}^d) + \alpha \mathcal{L}_{DRA} + \beta \mathcal{L}_{TRA} 
\label{eq:train}
\end{equation}
where $\alpha$, $\beta$ are parameters to control the effect of graph contrastive learning, and the first term is the negative conditional log-likelihood of the target token $y_t^d$.
%See appendix \ref{sec:data} for a detailed process of building the dataset.

\section{Experiments}
\textbf{Datasets}. To evaluate the effectiveness of our methods, we conduct experiments on SSN and PubMedCite datasets, as shown in Table~\ref{tab:stats}.
Following the previous method~\cite{an2021enhancing}, we have both transductive and inductive settings in PubMedCite.
For SSN, we use its original train/validation/test data splitting: 128,299/6,250/6,250 for inductive setting, and 128,400/6,123/6,276 for transductive setting.
Different from it, we keep the same train/validation/test set splitting: 178,100/7,036/7,608 in our PubMedCite in both settings.
This makes a more fair comparison of the influence of performance with different training setting, since we only have different citation correlations in inductive and transductive settings.
Following previous work, we use the abstracts of scientific documents as target summaries.

\textbf{Baselines}. We compare our method with: 1) extractive methods: LEAD~\cite{see2017get}, a simple method that selects the first n sentences; TextRank~\cite{mihalcea2004textrank}, a graph-based ranking method; TransformerEXT~\cite{liu2019text}, a transformer encoder-based method; BERTSUMEXT~\cite{liu2019text}, a BERT encoder based method;
BERTSUMEXT+~\cite{liu2019text}, a BERT encoder based method with 640 input tokens, 
2) abstractive methods: PTGEN+COV\\~\cite{see2017get}, a method based on copy mechanism;
TransformerABS~\cite{liu2019text}: a transformer encoder based method;
BERTSUMABS~\cite{liu2019text}, that uses the BERT encoder;
BERTSUMABS+~\cite{liu2019text}, the BERT encoder based method with 640 input tokens;
BART~\cite{lewis2020bart}, a competitive pre-trained language model for text generation;
CGSUM~\cite{an2021enhancing}, the SOTA method for the citation graph summarization task, based on the graph neural network.
%Moreover, we also provide the Oracle score as the upper bound, which is calculated based on the first 500 tokens of input documents in both datasets following previous methods. 
We report the F1 score of unigram overlapping(ROUGE-1), bigram overlapping(ROUGE-2), and longest common subsequence (ROUGE-L) between generated summaries and gold summaries~\cite{lin2003automatic}.

\textbf{Implementation Detail}. Our method is implemented by Python and Pytorch\footnote{Our model and data can be accessed at here \url{https://github.com/zhehengluoK/CitationSum}}.
We use the implementation of BERT, BART and PubMedBERT from Huggingface\footnote{ Models are initialized from checkpoints "bert-base-uncased", "microsoft/BiomedNLP-PubMedBERT-base-uncased-abstract", and "facebook/bart-base"}.
We run our experiments on multiple GPUs with 32G memory.
Following the previous method~\cite{liu2019text,xie2022pre}, we set the different learning rates to PLMs-based encoder and transformer based decoder.
We set the learning rate of the encoder to $2e-3$, that of the decoder to $2e-1$, the drop-out rate of the decoder to $0.4$, the training steps to 200000, the warm-up steps of the encoder to 20000, that of the decoder to 10000, the maximize a token number of input documents to 1240, and that of each reference of input documents to 100.
$\alpha,\beta$ on graph contrastive loss controlling, is set to 1.
We save checkpoint at every 200 steps, and select the best checkpoint according to the validation.
Due to the memory limitation, we set the maximum number of neighbours of input documents to 16.
For each document, we take tokens from its references as the negative tokens.
During the content selection of the test data set, we use the introduction of the source document to select contents from its references.
neighbour references of the test may be from the training data in the transductive setting, while the dataset is split into totally independent train/validation/test sets in the inductive setting.
\begin{table*}[!hbt]
    \Description{Rouge results on different datasets under different settings}
    \centering

    \caption{ROUGE F1 results of different models on SSN and PubMedCite. The results of our model are under 5 times running. $\dagger$ means outperform the existing model with best performance significantly ($p < 0.05$). Part results are from~\cite{an2021enhancing}.}
    %\setlength\tabcolsep{3pt}
   %\resizebox{\textwidth}{36mm}{
    \begin{tabular}{c|c|c|c|c|c|c|c|c|c|c|c|c}
    \hline
    \bf{Datasets}  & \multicolumn{6}{c|}{\bf{SSN}} & \multicolumn{6}{c}{\bf{PubMedCite}}\\
    \hline
     \bf{Setting}  & \multicolumn{3}{c|}{\bf{Inductive}} & \multicolumn{3}{c|}{\bf{Transductive}} 
     & \multicolumn{3}{c|}{\bf{Inductive}} & \multicolumn{3}{c}{\bf{Transductive}}\\
    \hline
    Metrics&	R-1&	R-2&	R-L&	R-1&R-2&	R-L&	R-1&	R-2&	R-L&R-1&	R-2&	R-L\\
    \hline
     LEAD&28.29&5.99&24.84&28.30&6.87&24.93&28.06&6.38&25.59&29.27&6.75&26.59\\
    %ORACLE&50.62&23.15&45.50&50.12&23.31&45.29&49.93&24.36&46.86&49.40&24.24&46.52\\
    \hline
    TextRank&36.36&9.67&32.72&40.81&12.81&36.47&38.87&11.35&34.42&39.00&11.37&34.50\\
    TransformerEXT&43.14&13.68&38.65&41.45&13.02&37.20&38.46&11.96&35.68&38.41&11.59&35.80\\
    BERTSUMEXT&42.41&13.10&37.97&41.68&13.31&37.42&38.72&11.96&35.86&38.82&11.96&35.98\\
    BERTSUMEXT+&44.28&14.67&39.77&43.23&14.59&38.91&39.14&12.02&36.18&39.18&12.09&36.22\\
    \hline
    PTGEN+COV&42.84&13.28&37.59&39.46&12.06&35.72&-&-&-&-&-&-\\
    \textit{Concat Nbr.Summ}&43.05&13.53&37.97&40.12&12.58&35.94&-&-&-&-&-&-\\
    TransformerABS&37.78&9.59&34.21&36.58&10.19&33.13&34.41&10.59&31.80&33.36&10.37&31.11\\
    \textit{+Copy}&41.22&13.31&37.22&40.83&14.71&36.93&38.36&11.86&35.55&38.72&11.96&35.86\\
    BERTSUMABS&43.73&15.05&39.46&40.38&14.07&36.54&39.07&11.78&35.33&39.11&11.78&36.38\\
    BERTSUMABS+&43.73&15.05&39.46&41.92&15.09&37.79&39.15&11.38&35.58&39.22&12.28&36.31\\
    \textit{Concat Nbr.Summ}&43.45&14.89&39.27&41.11&14.50&37.16&39.13&11.81&35.84&39.38&11.88&36.01\\
    \hline
    CGSUM&44.28&14.75&39.76&43.45&14.71&38.89&40.52&12.10&37.07&39.91&11.89&36.78\\
    CitationSum+BART&-&-&-&-&-&-&39.37&11.46&35.98&39.43&11.52&36.05\\
     CitationSum+BERT&44.72&15.03&40.12&44.07&15.02&39.47&
    39.48&12.24&36.47&39.62&12.29&36.54\\
\textit{+PubMedBERT}&$\textbf{45.01}^\dagger$&$\textbf{15.18}^\dagger$&$\textbf{40.59}^\dagger$&
    $\textbf{44.26}^\dagger$&$\textbf{15.45}^\dagger$&$\textbf{39.59}^\dagger$&
    $\textbf{41.62}^\dagger$&$\textbf{13.29}^\dagger$&$\textbf{37.54}^\dagger$&
    $\textbf{41.76}^\dagger$&$\textbf{13.36}^\dagger$&$\textbf{37.69}^\dagger$
    \\
    \hline
    \end{tabular}
    %}
    
    \label{tab:results1}
\end{table*}
\subsection{Results Analysis}
\subsubsection{Main Results}
We first show the ROUGE F1 score of different methods in both datasets in Table \ref{tab:results1}.
We investigate both BERT~\cite{kenton2019bert} and PubMedBERT~\cite{gu2021domain} as the encoder in our method.
Our method with the PubMedBERT-based encoder (CitationSum + PubMedBERT) presents the best performance among all baselines on both datasets when evaluating R-1 (ROUGE-1) and R-2 (ROUGE-2) for informativeness and R-L (ROUGE-L) for fluency.
When compared with CGSUM which also incorporates citation graphs to enhance summarization generation, it outperforms CGSUM in both inductive and transductive settings.
Different from CGSUM which uses abstracts of references, our method incorporates high-quality information from the full contents of references under the guidance of its semantic similarity with the source papers.
This proves the advantage of our method to make better and full use of references and the structure information of the citation graph.
It is also demonstrated by the superior performance of our method using BERT encoder compared with all BERT based abstractive methods including BERTSUMABS+, BERTSUMABS and BERTSUMABS+\textit{Concat Nbr.Summ}.
This proves essential to capture salient information about references and varying semantic correlations between scientific scripts and their references.
Moreover, although CGSUM is based on the LSTM backbone model, it outperforms pre-trained language model-based methods such as BERTSUMABS and CitationSum + BART.
This proves that the performance improvement of our method is not attributed to using the pre-trained language model, but due to make better leveraging the information of the citation graph.

Although it has been proven that BART shows better performance than BERT in text summarization, we surprisingly find that CitationSum + BART doesn't have the advantage for domain-specific scientific papers when compared with the CitationSum + BERT.
We can also observe that CitationSum + BERT underperforms CitationSum + PubMedBERT in all datasets and CGSUM in PubMedCite due to the limited vocabulary of BERT.
Since there are many terminologies in documents of PubMedCite, the PubMedCite dataset is more challenging for BERT based methods compared with SSN.
During experiments, we find that when using BERT's original tokenizer, there would be so many unrecognized tokens in PubMedCite to be set as [UNK] that causes much information loss and leads to many meaningless [UNK] tokens being generated in summaries.
In contrast, on SSN our method with BERT encoder outperforms CGSUM and has no tokens unrecognized as [UNK], and CitationSum + PubMedBERT has limited improvement on the performance of SSN compared with CitationSum + BERT.
This proves that the PubMedCite dataset with high technical domain-specific papers is more challenging when compared with SSN.
%Our model with the PubMedBERT based encoder (CIRCLE+PubMedBERT) shows an improvement when compared with the BERT encoder based model (CIRCLE+BERT) especially in the PubMedCite dataset.
% Although CIRCLE+BERT outperforms BERT encoder based abstractive methods including BERTSUMABS+ and BERTSUMABS due to the contrastive loss, it has different
% performance at SSN and PubMedBRT when compared with CGSUM and CIRCLE+PubMedBERT.
% We can see that CIRCLE+BERT underperforms CGSUM in PubMedCite, and CIRCLE+PubMedBERT yields a significant improvement on the performance compared with CIRCLE+BERT.
% During experiments, we find that when using BERT as the encoder, there are many unrecognized tokens in PubMedCite that are set as [UNK].
% Since there are many terminologies in documents of PubMedCite that can not be covered by the vocabulary of BERT.
% It leads to much information losses, and many meaningless [UNK] tokens to be generated in summaries.
% In the contrast, CIRCLE+BERT has better performance than CGSUM in SSN, and CIRCLE+PubMedBERT has limited improvement on the performance of SSN compared with CIRCLE+BERT.
%This proves that the PubMedCite dataset with high technical domain-specific papers is more challenging when compared with SSN.

Moreover, our models and CGSUM both have superior performance to other models that ignore information of references including BERT and transformer-based extractive and abstractive methods, as well as traditional methods TextRank and PTGEN+COV.
We can notice that simply appending the content from reference papers (Concat Nbr.Summ) in baseline methods presents limited benefit or even yield worse performance.
For example, the BERTSUMABS+\textit{Concat Nbr.Summ} with the extra input tokens from abstracts of references seldom underperforms the BERTSUMABS+.
This may be because abstracts of references can not provide useful information or even introduce extra noise as we mentioned before in Section \ref{sec:intro}.
This indicates that although leveraging references is beneficial to better understand scientific papers, it is vital to distinguish between salient and non-salient information in references.
%Besides, our method considers the whole text of the references, rather than only abstracts that can be insufficient for summarization.
%Our model and CGSUM shows an improvement when compared with BERT based extractive (second part) and abstractive (third part) strong baselines.
%Although they can exploit local contextual information of words in the source document, they ignore the content of the reference papers and also their connections to the source paper.
%However, , which also demonstrates the importance to distinguish between salient and nonsalient information in the reference papers.
\begin{table*}[!hbt]

%   \resizebox{\textwidth}{15mm}{
    \Description{Ablation results table}
    \caption{ROUGE F1 results of our model under different settings on SSN and PubMedCite.} 
    \begin{tabular}{c|c|c|c|c|c|c|c|c|c|c|c|c}
    \hline
    \bf{Datasets}  & \multicolumn{6}{c|}{\bf{SSN}} & \multicolumn{6}{c}{\bf{PubMedCite}}\\
    \hline
     \bf{Setting}  & \multicolumn{3}{c|}{\bf{Inductive}} & \multicolumn{3}{c|}{\bf{Transductive}} 
     & \multicolumn{3}{c|}{\bf{Inductive}} & \multicolumn{3}{c}{\bf{Transductive}}\\
    \hline
    Metrics&	R-1&	R-2&	R-L&	R-1&R-2&	R-L&	R-1&	R-2&	R-L&R-1&	R-2&	R-L\\
    \hline
    W/O Contra&44.77&14.70&39.66&43.92&14.83&38.84&40.16&12.51&37.13&39.99&12.43&36.93\\
    W/O DRA&44.83&14.91&40.02&44.07&14.88&38.97&40.17&12.58&37.18&40.07&12.52&37.01\\
    W/O TRA&44.94&14.95&40.18&44.15&14.91&39.09&40.90&12.93&37.16&40.08&13.17&37.25\\
    W/O Concate&44.96&15.13&40.57&44.22&15.53&39.37&41.38&13.29&37.51&41.42&13.29&37.69\\
    Full&$\textbf{45.01}^\dagger$&$\textbf{15.18}^\dagger$&$\textbf{40.59}^\dagger$&
    $\textbf{44.26}^\dagger$&$\textbf{15.45}^\dagger$&$\textbf{39.59}^\dagger$&
    $\textbf{41.62}^\dagger$&$\textbf{13.29}^\dagger$&$\textbf{37.54}^\dagger$&
    $\textbf{41.76}^\dagger$&$\textbf{13.36}^\dagger$&$\textbf{37.69}^\dagger$
    \\
    \hline
    \end{tabular}
 %   }
    
    \label{tab:abla}
\end{table*}

\subsection{Ablation Study}
To further clarify the contribution of each component in our method, we perform experiments on our method and several ablations which respectively removes contrastive learning (W/O Contra), document representation alignment (W/O DRA), token representation alignment (W/O TRA), and concatenation of token representations of references (W/O Concate).
As shown in Table \ref{tab:abla}, the performance of our method suffers least without concatenation, indicating the efficacy of our hierarchical graph contrastive learning in delivering salient information from references to encoding source documents. 
It can also be demonstrated by the lowest ROUGE score of W/O Contra since without our design of contrastive learning, the model can be faced with difficulty finding useful messages from concatenated reference content.
Moreover, the drops brought by W/O DRA and W/O TRA separately manifest the essential of inter-document and inner-document connections in our design. 
In the absence of either of them, information propagation in the heterogeneous graph would be blocked, inhabiting the model's ability to fully understand the source paper.
\subsection{Parameter Impacts}
We further explore the influence of parameter $\rho$ of controlling edge weights on the performance.
$\rho$ is used to filter edges between the source document and its references in the citation graph, according to their semantic similarity as shown in Section \ref{sub:hgc}. 
From Table \ref{tab:eff}, we can see that our model yields the best performance when $\rho=0.7$.
A too low value of $\rho$ ($\rho < 0.7$) could incorporate references that have low semantic similarity with the source document.
Thus extra noise could be introduced.
In contrast, the too high value of $\rho$ ($0.7<\rho$) could filter references with useful information. 
\begin{table}[!hbt]
\small
    \Description{ROUGE F1 results of our model with different values of $\rho$ under inductive setting}
    \centering
    \small
    \caption{ROUGE F1 results of our model with different values of $\rho$ under inductive setting, that controls weights of edges in the citation graph.}
    %\resizebox{0.38\textwidth}{9mm}{
    \begin{tabular}{c|c|c|c|c|c|c}
    \hline
    \bf{Datasets}  & \multicolumn{3}{c|}{\bf{SSN}} & \multicolumn{3}{c}{\bf{PubMedCite}}\\
    \hline
    Metrics&	R-1&	R-2&	R-L&R-1&R-2&R-L\\
    \hline
    0.5&44.69&14.72&40.97&41.09&12.54&37.04\\
    0.6&44.94&14.98&40.02&41.20&13.29&37.13\\
    0.7&$\textbf{45.01}$&$\textbf{15.18}$&$\textbf{40.59}$&
    %$\textbf{44.26}$&$\textbf{15.45}$&$\textbf{39.59}$&
    $\textbf{41.62}$&$\textbf{13.29}$&$\textbf{37.54}$\\
    %$\textbf{41.76}$&$\textbf{13.36}$&$\textbf{37.69}$\\
    0.8&44.82&14.88&40.38 &41.27&12.95&37.18\\
    \hline
    \end{tabular}
    
    \label{tab:eff}
\end{table}
% As shown in Table \ref{tab:abla}, the performance of our method suffers most without concatenation, which ignore all information from references.
% However, our method still underperforms CGSUM without contrastive learning, since the abundant information in the content of references can be noisy to the understanding of the source document.
% Therefore, our method with DRA presents competitive or even better results compared with CGSUM.
% Without TRA, it can be deemed as an solely implicit reconstruction of the citation graph, which allows our method to exploit the connections between the whole content of source document and its references.
% Yet, similar to our method with only TRA, it is a partial model of the hierarchical semantic correlations between source document, its references and their tokens, which blocks the interactions among inner-document and intra-document and inhabits the ability to fully understand the source paper.
\subsection{Case Study}
In Appendix \ref{appd: example} Table \ref{table:case}, we show the generated summary by our method of an example document along with its reference.
Tokens that are semantically correlated to the example document, are marked with blue colour.
It shows that our model is able to generate a coherent summary for the document that is highly semantically related to its gold summary.
Compared with the abstract of its reference, the selected high-quality content of its reference can provide more useful information indicated by a higher mean rouge score with the example document.
This is also illustrated in the selected content of references which have a large number of related tokens that are marked with blue colour.
Our method can effectively recognize salient information from references for fully exploiting the hierarchical connections among documents and tokens, which is utilized to generate concise summaries with many relevant tokens marked in blue colour.
\section{Conclusion}
We propose a novel self-supervised framework for the summarization of scientific papers based on the citation graph and contrastive learning, to make better use of references and citation correlations.
We also propose a novel biomedical domain-specific dataset, by which we expect to support evaluation and method development in the research community.
Experimental results on two benchmark datasets show the effectiveness of our proposed method.
There are several limitations that could be addressed in the future:
1) For the high-quality content selection of references, we only utilize the ROUGE score as the semantic similarity metric and evaluation metric. 
Other advanced metrics such as the BERTScore can be explored in the future.
2) The PLM encoder in our model only encodes a limited number of input tokens. However, the average document length of documents is up to four thousand. It is expected to address this issue in the future.
3) We only utilize the document contents of source documents and their references.
There is rich structural information for scientific documents, such as title, introduction, and related work, that can be incorporated further. 

\begin{acks}
We would like to thank the Computational Shared Facility at The University of Manchester for offering computing resources. We also want to thank Jimin Huang and Paul Thompson for their suggestions and feedback on this paper.
This research is supported by the Biotechnology and Biological Sciences Research Council (BBSRC), BB/P025684/1, the Joint Laboratory on Credit Technology, the National Key Research and Development Program of China (No.2021ZD0113304), and Application Foundation Frontier Project of Wuhan (Grant NO.2020010601012168). 
\end{acks}
%Directly utilizing full contents of references instead of abstracts can be problematic for introducing abundant information.

\bibliographystyle{ACM-Reference-Format}
\bibliography{custom}

\appendix
\section{Summarisation Examples}
\label{appd: example}
\begin{table*}[hbt!]
\small
\Description{An example of generated summary by our method on the PubMedCite dataset.
The gold summary is the abstract of the document.}
\caption{An example of generated summary by our method on the PubMedCite dataset.
The gold summary is the abstract of the document.}
\begin{tabular}{p{0.9\textwidth}}
\hline \textbf{Gold:} 
Recent work has provided new insights into the \textcolor{blue}{mechanism} of \textcolor{blue}{spindle assembly}.
Growing evidence supports a model in which the \textcolor{blue}{small GTPase Ran} plays a central \textcolor{blue}{role} in this process.
Here, we examine the evidence for the existence of a \textcolor{blue}{RanGTP gradient} around \textcolor{blue}{mitotic chromosomes} and some controversial data on the role that \textcolor{blue}{chromosomes} play in \textcolor{blue}{spindle assembly}.
We review the current knowledge on the \textcolor{blue}{Ran} downstream targets for \textcolor{blue}{spindle assembly} and we focus on the multiple roles of TPX2, one of the targets of \textcolor{blue}{RanGTP} during \textcolor{blue}{cell} division.
\\
\hline
\textbf{Generated:} 
Recent studies have shown that the targeting protein for xklp2 ( tpx2 ), one of the central targets of \textcolor{blue}{ran} in \textcolor{blue}{spindle assembly}, is a central target of \textcolor{blue}{ran}.
In this review , we examine some controversial data on the \textcolor{blue}{role} of \textcolor{blue}{chromosomes} and the ran model in \textcolor{blue}{spindle} formation.
We then review our current knowledge on the downstream targets of rangtp during mitosis and we focus on the multiple functions performed by tpx2 in mitosis.
Finally, we discuss the functions of tpx2 in the regulation of \textcolor{blue}{spindle assembly} and discuss the potential of tpx2 as a target for \textcolor{blue}{ran} in mitosis and in other \textcolor{blue}{cell} cycle regulated processes.
\\
\hline
\textbf{Selected content of reference:} 
While this is an attractive model, it is likely that the true situation is more complicated and that the location of other \textcolor{blue}{Ran} pathway components can play a \textcolor{blue}{role} in regulating \textcolor{blue}{Ran-GTP} distribution.
Ran is a \textcolor{blue}{GTPase} that is required for nuclear transport, \textcolor{blue}{cell} cycle control, \textcolor{blue}{mitotic} \textcolor{blue}{spindle formation}, and postmitotic nuclear \textcolor{blue}{assembly,} although the \textcolor{blue}{mechanism} of this targeting has not been reported.
We wished to determine whether SUMO-1 plays a \textcolor{blue}{role} in RanGAP1's targeting to the \textcolor{blue}{spindle} during mitosis.
We wished to determine more precisely where RanGAP1 localized on the \textcolor{blue}{spindle}.
The \textcolor{blue}{small GTPase Ran} plays a key role in diverse cellular \textcolor{blue}{functions} including nucleocytoplasmic transport, nuclear envelope formation, and \textcolor{blue}{spindle assembly}. 
This unique localization of the \textcolor{blue}{Ran} regulators strongly suggests that there is a \textcolor{blue}{RanGTP} concentration \textcolor{blue}{gradient} across the interphase nuclear envelope and on the condensed \textcolor{blue}{mitotic chromosomes}.
(\textbf{Mean ROUGE: 0.519})
\\\hline
\textbf{Abstract of reference:} 
The microtubule cytoskeleton plays a pivotal \textcolor{blue}{role} in cytoplasmic organization, \textcolor{blue}{cell} division, and the correct transmission of genetic information. 
In a screen designed to identify fission yeast genes required for chromosome segregation, we identified a strain that carries a point mutation in the SpRan \textcolor{blue}{GTPase}. 
\textcolor{blue}{Ran} is an evolutionarily conserved eukaryotic \textcolor{blue}{GTPase} that directly participates in nucleocytoplasmic transport and whose loss affects many biological processes.
Recently a transport-independent effect of \textcolor{blue}{Ran} on \textcolor{blue}{spindle} formation in vitro was demonstrated, but the in vivo relevance of these findings was unclear. 
Here, we report the characterization of a Schizosaccharomyces pombe \textcolor{blue}{Ran GTPase} $\cdots$ (\textbf{Mean ROUGE: 0.241}) 
%function mutant in which nucleocytoplasmic protein transport is normal, but the microtubule cytoskeleton is defective, resulting in chromosome missegregation and abnormal cell shape. (\textbf{Mean ROUGE: 0.241}) 
%These abnormalities are exacerbated by microtubule destabilizing drugs, by loss of the spindle checkpoint protein Mph1p, and by mutations in the spindle pole body component Cut11p, indicating that SpRan influences microtubule integrity. As the SpRan mutant phenotype can be partially suppressed by the presence of extra Mal3p, we suggest that SpRan plays a role in microtubule stability.
\\\hline
\end{tabular}

\label{table:case}
\end{table*}

\end{document}